\documentclass[aps,prl,twocolumn,showpacs,a4paper,groupedaddress]{revtex4}

\usepackage{graphicx,amsmath,SIunits}
\usepackage[latin1]{inputenc}
\usepackage{latexsym,stmaryrd}

\begin{document}
\title{Controlling Anderson localization in disordered photonic crystal waveguides}

\author{P. D. Garc\'{i}a}
\email{dgar@fotonik.dtu.dk}
\author{S. Smolka}
\author{S. Stobbe}
\author{P. Lodahl}
\email{pelo@fotonik.dtu.dk}
\homepage{http://www.fotonik.dtu.dk/quantumphotonics}
\affiliation{DTU Fotonik, Department of Photonics Engineering,\
Technical University of Denmark,\ Ørsteds Plads 343,\ DK 2800
Kgs.\ Lyngby,\ Denmark}

\date{\today}

\begin{abstract}

We prove Anderson localization in a disordered photonic crystal
waveguide by measuring the ensemble-averaged localization length,
$\xi$, which is controlled by the dispersion of the photonic
crystal waveguide. In such structures, $\xi$ shows a 10-fold
variation between the fast- and the slow-light regime and, in the
latter case, it becomes shorter than the sample length thus giving
rise to strongly confined modes. The dispersive behavior of $\xi$
demonstrates the close relation between Anderson localization and
the photon density of states in disordered photonic crystals,
which opens a promising route to controlling and exploiting
Anderson localization for efficient light confinement.

\end{abstract}

 \pacs{(42.25.Dd, 42.25.Fx, 46.65.+g, 42.70.Qs)}

\maketitle

Quantum optics and quantum information technologies require
enhancement of light-matter interaction by, for example, confining
light in a highly engineered nanocavity \cite{Noda}.\ Quite
remarkably, an alternative route towards light confinement
exploits multiple scattering of light in disordered photonic
structures, as originally proposed for electrons by P. W. Anderson
\cite{Anderson}.\ The mechanism responsible for Anderson
localization is wave interference and hence occurs not only for
electrons but also for, e.g., microwaves \cite{Chabanov}, acoustic
waves \cite{Tigelen}, and even Bose-Einstein condensated matter
waves \cite{Bose} thus illuminating the multidisciplinarity of the
research field.\ In the case of light, indications of three-dimensional (3D)
Anderson localization have been observed in random dielectric
materials like powders composed of particles with casual shapes
and sizes \cite{powder}.\ In these systems no control can be
exerted over the frequency or spatial extent of the localized
modes.

A promising proposal on how to control multiple scattering is to
introduce disorder in photonic crystals \cite{John}: the
interference of multiply scattered light is expected to form
Anderson-localized modes that appear randomly in space but at
frequencies in or near the photonic crystal band gap.\ The
characteristic length scale of Anderson localization is the
localization length $\xi$, which is the exponential decay length
of the confined modes after averaging over many realizations of
disorder. In one- and two-dimensional (1D, 2D) systems,
localization occurs for any degree of disorder when the sample
length exceeds $\xi$ \cite{mott}, in which case the photonic
conductor becomes an insulator.\ Contrary to non-dispersive
systems, photonic crystals offer the possibility to alter the
photon density of states (DOS) allowing to control macroscopic
transport properties \cite{DOS} or spontaneous emission of photons
\cite{Lodahl}, and in particular to achieve dispersive Anderson
localization.

\begin{figure}[h!]
    \begin{center}
   \includegraphics[width=6cm]{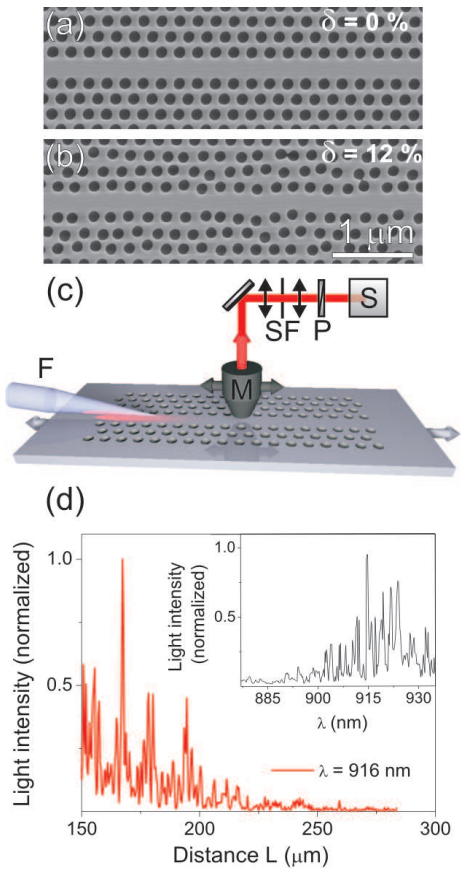}

    \caption{ \label{1} (Color online) (a) and (b) Scanning electron microscope images of photonic crystal
    waveguides (top-view) with engineered disorder on the positions of the holes of $\delta = 0\%$ and $\delta = 12\%$ (standard deviation relative to the lattice constant). (c) Setup used to measure the localization length.\ The tip of a single mode tapered fiber (F) is placed on the
    photonic crystal waveguide.\ TE-polarized light scattered from the waveguide is collected with an optical microscope objective (M), a spatial filter (SF), and a polarizer (P) and
    sent to a spectrometer (S). (d) Measurement of strongly fluctuating light intensity in the slow-light regime of a PCW with $\delta = 0\%$ versus the distance from the light source.\ The inset of the
    figure shows a spectrum taken at $L = 150\,\micro\text{m}$ from the fiber tip.}
    \end{center}
    \end{figure}

In this Letter, we show how to control and accurately tune
Anderson localization of light to a spectral range of particular
interest using dispersion in photonic crystal.\ In particular, we
use a 1D disordered photonic crystal waveguide (PCW) to prove the
close relation between Anderson localization of light and the
DOS.\ Indeed, photonic crystals owe their success to the fact that
their DOS can be accurately tailored \textit{a priori} and we show
how this property can be exploited to tailor Anderson
localization.\ By explicitly measuring the dispersive localization
length, we unambiguously demonstrate that the strongly confined
modes appearing in the slow-light regime of PCWs \cite{Vollmer}
are due to 1D Anderson localization, which has been questioned
recently \cite{Hughes}.\ In addition, we interpret our
experimental data on the wavelength-dependent localization length
with a model for the DOS of a PCW thereby explicitly linking the
localization length and the DOS.

Our samples consist of a membrane with a high refractive index
material (GaAs, $n=3.54$) in which light is confined by total
internal reflection.\ An ordered lattice of holes is etched in the
structure forming a 2D photonic band gap that suppresses in-plane
propagation of light.\ A waveguide is engineered in the structure
by leaving out a row of holes.\ Light propagation in an ideal PCW
is described by Bloch modes with a dispersion relation
$\omega(k)$, where $\omega$ is the wave frequency and $k$ is the
wave vector.\ PCWs can generate slow light, i.e., light with a
very low group velocity, $v_\text{g} =\partial \omega/\partial k
=c/n_\text{g}$, where c is the speed of light in vacuum.\ The
slow-down factor $n_\text{g}$, also referred to as the group
index, is directly proportional to the DOS per unit length of the
propagating mode in the PCW: $\text{DOS}=(1/\pi) \partial
k/\partial \omega = n_\text{g}/(\pi c)$.\ Imperfections in a PCW
lead to multiple scattering of light and the dispersion relation
breaks down \cite{Thomas} thus inducing localized modes in the
slow-light regime where the sample length exceeds $\xi$.

To investigate the impact of disorder on the light propagation in
PCW we randomly vary the hole positions in the three rows above
and below the waveguide using a Gaussian random number generator
function (Box-Muller).\ The degree of disorder in each sample,
$\delta$, is characterized by the standard deviation of the hole
position with respect to the lattice constant varying from $\delta
= 0\%$ to $\delta = 12\%$ (Fig.~\ref{1}(a), (b)).\ The samples
consist of a triangular lattice of holes with a lattice constant
$a = 240\,\text{nm}$, a filling fraction $f = 0.330\pm0.006$,
membrane height $h=160\pm5\,\text{nm}$, and a total length
$L_0=1\,\text{mm}$.\

\begin{figure}[h!]
    \begin{center}
   \includegraphics[width=8cm]{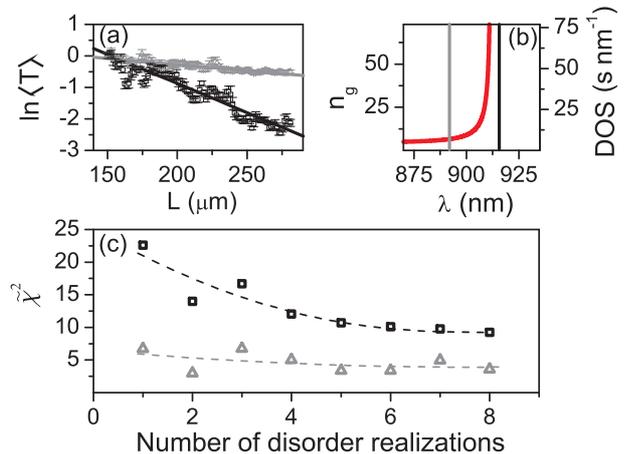}
    \caption{ \label{2} (Color online) (a) Exponential fit of the ensemble-averaged light transmission in PCWs with $\delta = 0\%$ as a function of the distance from the source for two
    wavelengths: $\lambda_1 = 890\,\text{nm}$ (gray points) and $\lambda_2 = 916\,\text{nm}$ (black points).\ The spectral resolution is $2\,\text{nm}$ and
    the slope equals the inverse of the localization
    length.\ (b) The calculated density of states (DOS) and group index ($n_\text{g}$) of an ideal structure without disorder. The two specific wavelengths $\lambda_1$ (gray line) and $\lambda_2$ (black line) are indicated that lie in the fast- and slow-light regimes,
    respectively.\ The calculated spectral position of the cutoff of the
    ideal waveguide mode gives the best fit to our experimental data (explained in the text).\ (c) Plot of the reduced chi-square, $\widetilde{\chi}^2$, for the exponential fit to the ensemble-averaged light transmission in the fast- (gray triangles)
and slow-light regime (black squares) versus number of
realizations of disorder.\ The dashed lines are guides to the
eyes.}
    \end{center}
    \end{figure}

To characterize Anderson localization in PCWs we use the optical
setup illustrated in Fig.~\ref{1}(c).\ A continuous wave
Ti:sapphire laser tuneable  within $\lambda=700-1000\,\text{nm}$
is coupled into a single mode tapered fiber with a tip diameter
comparable to the waveguide width.\ The fiber evanescent mode
couples to the waveguide mode by placing the fiber tip in close
proximity of the PCW.\ We measure light scattered out-of-plane
from the PCW with a high amplification microscope objective (NA =
0.8) as a function of the wavelength and the distance $L$ from the
light source, i.e., the fiber tip.\ The measurement starts at $L =
150\,\micro\text{m}$ from the fiber tip to avoid any spurious
effect due to the evanescent field or light that is not coupled to
the waveguide mode.\ Fig.~\ref{1}(d) shows a measurement of
scattered light intensity versus $L$ in the slow-light regime (at
$\lambda = 916\,\text{nm}$ where $n_\text{g}
> 15$) for a single realization of disorder.\ In this case, $\delta =
0\%$, the PCW is only affected by intrinsic unavoidable disorder
introduced in the fabrication process.\ The inset of the figure
shows a spectrum of the scattered light intensity at $L =
150\,\micro\text{m}$.\ The strong fluctuation in the light
intensity is a signature of 1D Anderson-localized modes.\ The
modes appear to be spatially and spectrally separated, which
constitutes a criterion for Anderson localization \cite{thouless}.

\begin{figure}[t]
    \begin{center}
   \includegraphics[width=8.5cm]{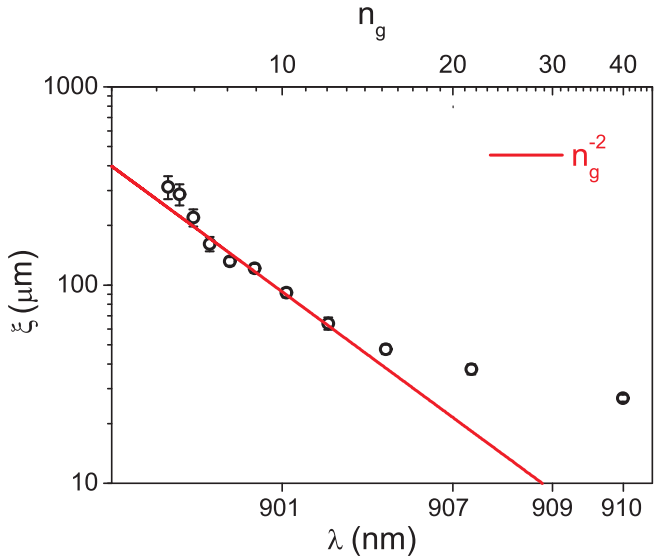}
    \caption{ \label{3} (Color online) Plot of the localization length, $\xi$, of the PCW with $\delta = 0\%$ as a function of the wavelength and the group index of the ideal structure $n_\text{g}$.\
    A strongly dispersive localization length is found that reaches a minimum in the slow-light regime ($n_\text{g}>15$).\ The scaling $\xi \propto n_\text{g}^{-2}$ (red curve) is obtained from a model
    of the light scattering cross section in
    the PCW.}
    \end{center}
    \end{figure}
In the Anderson localization regime the fluctuating light
transmission, $T(L)$, decays exponentially after
ensemble-averaging, giving $\text{ln}\langle T \rangle = - L/2\xi$
\cite{Beenakker}.\ We have used this property to extract the
localization length and confirm the Anderson localization
criterion of $\xi < L$.\ To do this, we perform ensemble averaging
by measuring the light leakage versus $L$ and wavelength for $8$
different realizations of disorder.\ In particular, we probe
different spatial realizations of disorder by moving the fiber
together with the microscope objective along the waveguide and
repeating the measurement.\ Fig.~\ref{2}(a) shows the linear fit
of $\text{ln}\langle T \rangle $ in the sample with $\delta = 0\%$
in the fast-light regime $n_\text{g}(\lambda_1)<15$ and slow-light
regime $n_\text{g}(\lambda_2)>15$ (corresponding to low and high
DOS, respectively, see Fig.~\ref{2}(b)).\ For these spectral
positions we extract the localization lengths $\xi(\lambda_1) =
(263 \pm 28)\,\micro\text{m}$ and $\xi(\lambda_2) = (27 \pm 2)\
\micro\text{m}$.\ We obtain $n_\text{g}$ from numerical
simulations of the disorder-free ideal structure, which quantifies
the spectral position with respect to the cutoff of the ideal
waveguide mode.\ This spectral position is obtained from the best
fit to our experimental data with a 1D model (explained later
on).\ It is within the uncertainty in the waveguide mode band edge
position, which is caused by uncertainties in the fabrication
parameters given previously.

Fig.~\ref{2}(c) shows the goodness of the fits,
$\widetilde{\chi}^2$, performed in Fig.~\ref{2}(a) to test the
degree of convergence of the ensemble average to a
single-exponential decay.\ The fluctuations in the data are due to
speckles not fully ensemble-averaged.\ A disorder realization in
our experiment consists of a scan of $T(L)$ varying $L$ over $130\
\micro \text{m}$.\ Thus, the number of different disorder
realizations that we can perform without adding repeated
statistics to the ensemble average is limited by the sample
length, $L_0 = 1\ \text{mm}$.\ Furthermore, the goodness of the
fit does not reach the optimum value $\widetilde{\chi}^2=1$
corresponding to a single-exponential decay, which may be due to
different reasons.\ Deviations from perfect verticality of the air
holes can break the symmetry in the out-of-plane direction and
couple TE- to TM-polarized waveguide modes giving rise to strongly
localized modes \cite{Topolancik}.\ This polarization mixing
mechanism may lead to multi-exponential decay of the
ensemble-averaged transmission that is not resolved in the present
experiment.

Fig.~\ref{3} contains the main contribution of this Letter: it
shows the dispersive behavior of the measured localization
length.\ We observe a 10-fold variation in the localization length
with wavelength.\ In the slow-light regime, $\xi$ reaches its
minimum value of $\sim (27 \pm 2)\,\micro\text{m}$ and becomes
much smaller than $L_0$, giving rise to strongly
Anderson-localized modes.\ The dispersive behavior of the
localization length gives directly control over the extension of
the modes, which can be precisely tuned by varying the wavelength.
We can also tune the waveguide mode by varying the fabrication
parameters (typically $a$ and $f$) thus, controlling the frequency
of localized modes.

Any loss mechanisms of the light trapped in the PCW will in
general influence the measured localization length depending on
the spatial length scale of the loss process.\ In presence of
light losses, we have $\text{ln}\langle T \rangle = - L/2\ell$
where the decay length $\ell$ is defined as ${\ell}^{-1} =
{\xi}^{-1} + {\ell_i}^{-1} + {\ell_e}^{-1}$, where $\ell_i$ is the
material inelastic absorption length and $\ell_e$ is the
extinction length associated with out-of-plane losses.\ GaAs has a
very low optical absorption coefficient ($ <100\,\text{cm}^{-1}$
at a wavelength of $\lambda=915\,\text{nm}$) corresponding to
$\ell_i > 1\,\text{m}$.\ This value might be reduced by surface
effects at the holes of the photonic crystal but is still expected
to be much larger than $L_0$.\ Furthermore, we quantify $\ell_e$
applying our fabrication parameters to recent 3D numerical
simulations of Bloch-mode scattering in PCWs \cite{Lalanne}.\ In
particular, for $\delta = 0\%$, for which the standard deviation
of the hole positions is smaller than 2 nm, we obtain $\ell_e \sim
400\,\text{mm}$. Both $\ell_i$ and $\ell_e$ are much larger than
$L_0$, thus not affecting the localization length we extract from
our data, i.e., we can approximate ${\ell} \cong {\xi}$ at least
for weak disorder.

In the following, we model the function $\xi(n_\text{g})$.\ The
relation $\xi \approx N\cdot \ell_\text{s}$ holds for 1D systems
\cite{Beenakker,Saenz}, where $N$ is the number of electromagnetic
modes that the system can sustain and $\ell_\text{s}$ is the
scattering mean free path.\ Our samples are single-mode PCWs and
hence the localization length is $\xi \approx \ell_\text{s}$.\ The
dispersive behavior of $\ell_\text{s}$ in 3D photonic crystals has
been measured and explained in terms of the DOS very recently
\cite{PRBscattering}.\ Two separate mechanisms determine
$\ell_\text{s}$: the excitation of the scatterer and the
subsequent radiation from the scatterer \cite{Froufe}.\ The
coupling to the scatterer is described by the DOS of the mode of
the excitation beam \cite{Mcphedran}, which in our system is
proportional to the group index of the waveguide mode.\ The second
process is described by the local density of states (LDOS).\
Ignoring the minor contributions of coupling to unconfined
radiation modes, the LDOS is also determined by the group index of
the waveguide \cite{Hughes_purcell,Toke}.\ This applies to every
scattering event giving rise to a modified scattering cross
section $\sigma$ in PCWs scaling as $\sigma\propto {n_\text{g}}^2
(\omega) $.\ The scattering mean free path in a random medium can
be expressed as $\ell_\text{s} = 1/\rho_\text{s} \sigma$
\cite{Sheng}, where $\rho_\text{s}$ is the density of scatterers.\
For 1D single-mode PCWs we therefore predict $\xi \propto
{n_g}^{-2} (\omega)$, which is in very good agreement with our
experimental data in the fast-light regime. The red curve in
Fig.~\ref{3} represents the best fit to our experimental results.\
It gives the spectral position of the waveguide mode cutoff used
in Fig.~\ref{2}(b).\ The fitting parameters are the number of data
fitted and the spectral position of the cutoff of the waveguide
mode.\ The same scaling of $\xi \propto {n_g}^{-2}$ can also be
recovered from 1D random matrix theory \cite{FroufeRM} and it is
confirmed by 3D numerical simulations in PCWs \cite{Lalanne}.\ Our
model explains not only the dispersion in the localization length
but also the scaling of light losses with group index in PCWs as
${v_\text{g}}^{-2}$ (${n_\text{g}}^2$).\ Such a scaling has been
systematically observed in the literature \cite{losses} and our
model provides a physical explanation.\ From the data in
Fig.~\ref{3} we observe that the model breaks down deep in the
slow-light regime where strongly localized Anderson modes appear.\
This is expected since the model is based on the calculated DOS of
the ideal structure without disorder, which is modified in the
regime of strong multiple scattering and Anderson localization.

\begin{figure}[t]
    \begin{center}
   \includegraphics[width=7.5cm]{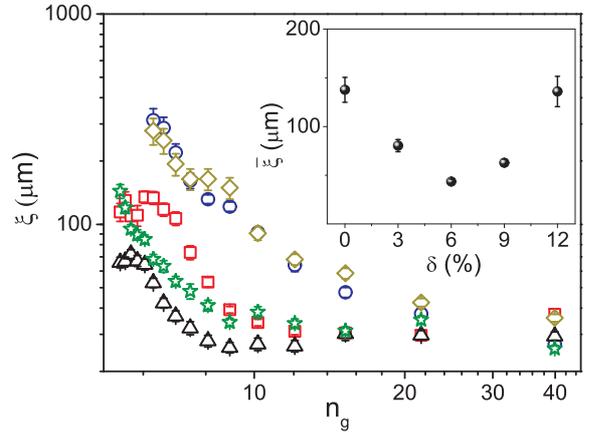}
    \caption{ \label{4} (Color online) Localization length, $\xi$, as a function of the group index, $n_\text{g}$, measured in PCWs with various amounts of disorder $\delta = 0\%$ ($\circ$), $3\%$
($\boxempty$), $6\%$ ($\triangle$), $9\%$ ($\star$), and $12\%$
($\diamond$).\ The inset shows the mean value of $\xi$ as a
function of $\delta$.}
    \end{center}
    \end{figure}

Finally, Fig.~\ref{4} plots the measurement of the localization
length in samples with increasing amount of disorder and the mean
value of the localization length, $\overline \xi$.\ The latter
decreases with disorder, reaches a minimum for $\delta = 6\%,$ and
increases for $\delta> 6\%$.\ This behavior is a clear proof that
losses are not dominant in our experiment even for strong disorder
where $\ell_e$ is predicted to shorten \cite{Lalanne}.\ The
increase of $\overline \xi$ for $\delta> 6\%$ cannot be explained
by an increase of light losses (decrease of $\ell_e$).\ That would
lead to a decrease of the measured decay length as opposed to our
observations.\ The behavior of $\overline \xi$ for weak disorder
is predicted in Ref. \cite{Lalanne} and we propose here a possible
explanation for the increase observed for large amounts of
disorder.\ Thus, the local disorder introduced in our samples only
in the three rows above and below the waveguide could imply that
several spatial modes, $N$, are effectively introduced when
increasing disorder.\ This would increase the localization length
according to $\overline \xi\propto N\cdot \ell_\text{s}$
effectively decoupling $\xi$ and $\ell_\text{s}$ and therefore
enable a regime of light diffusion where $\ell_\text{s}<L<\xi.$
This could open a new route to investigate the crossover between
diffusion and localization regimes in a dispersive quasi-1D
system, which is not possible in standard quasi-1D disordered
systems \cite{Saenz,bertolotti}.

In conclusion, we have demonstrated the close relation between
Anderson localization of light and the electromagnetic DOS in
PCWs.\ We explain multiple light scattering in PCWs with a DOS
dependent scattering cross section explaining the strong
dispersion of the localization length and the appearance of
strongly confined modes in the slow-light regime.\ These results
are of fundamental importance since they impose limitations for
slow-light devices based on PCWs such as single-photon sources
\cite{Toke}.\ At the contrary the strongly confined
Anderson-localized cavities with tailored properties appear very
appealing candidates for experiments on cavity quantum
electrodynamics \cite{Sapienza} or random lasing.

%\section{ACKNOWLEDGMENTS}

We thank Johan Raunkjær Ott for fruitful discussions.\ We
gratefully acknowledge the Council for Independent Research (Technology and
Production Sciences and Natural Sciences) for financial
support.

%--------------------------------------------------------------------

\end{document}